\documentclass[sigconf,balance]{acmart}

\usepackage{enumitem}
\usepackage{graphicx}
\usepackage{xcolor}
\usepackage{algorithm}
\usepackage{algpseudocode}

\usepackage{multirow}
\usepackage{tabularx, array}
\usepackage{cellspace}
\addparagraphcolumntypes{X} 
\newcolumntype{C}[1]{>{\centering\let\newline\\\arraybackslash\hspace{0pt}}m{#1}}

\usepackage{makecell}  

\newcolumntype{C}[1]{>{\centering\arraybackslash}m{#1}}
\newcolumntype{L}[1]{>{\arraybackslash}m{#1}}

\usepackage{caption}
\copyrightyear{2024}
\acmYear{2024}
\setcopyright{rightsretained}
\acmConference[SIGSIM PADS '24]{38th ACM SIGSIM Conference on Principles of Advanced Discrete Simulation}{June 24--26, 2024}{Atlanta, GA, USA}
\acmBooktitle{38th ACM SIGSIM Conference on Principles of Advanced Discrete Simulation (SIGSIM PADS '24), June 24--26, 2024, Atlanta, GA, USA}
\acmDOI{10.1145/3615979.3662149}
\acmISBN{979-8-4007-0363-8/24/06}

\begin{document}

\title{Loyal Wingman Assessment: Social Navigation for Human-Autonomous Collaboration in Simulated Air Combat}

\author{Joao~P.~A.~Dantas}
\email{dantasjpad@fab.mil.br}
\orcid{0000-0003-0300-8027} 
\affiliation{%
  \institution{Institute for Advanced Studies}
  \city{Sao Jose dos Campos}
  \country{Brazil}
}

\author{Marcos~R.~O.~A.~Maximo}
\email{mmaximo@ita.br}
\orcid{0000-0003-2944-4476} 
\affiliation{%
  \institution{Aeronautics Institute of Technology}
  \city{Sao Jose dos Campos}
  \country{Brazil}
}

\author{Takashi~Yoneyama}
\email{takashi@ita.br}
\orcid{0000-0001-5375-1076} 
\affiliation{%
  \institution{Aeronautics Institute of Technology}
  \city{Sao Jose dos Campos}
  \country{Brazil}
}

\begin{abstract}

This study proposes social navigation metrics for autonomous agents in air combat, aiming to facilitate their smooth integration into pilot formations. The absence of such metrics poses challenges to safety and effectiveness in mixed human-autonomous teams. The proposed metrics prioritize naturalness and comfort. We suggest validating them through a user study involving military pilots in simulated air combat scenarios alongside autonomous loyal wingmen. The experiment will involve setting up simulations, designing scenarios, and evaluating performance using feedback from questionnaires and data analysis. These metrics aim to enhance the operational performance of autonomous loyal wingmen, thereby contributing to safer and more strategic air combat.~\href{https://github.com/jpadantas/social-navigation-metrics}{[Code]}\footnote{\href{https://github.com/jpadantas/social-navigation-metrics}{Code: https://github.com/jpadantas/social-navigation-metrics}}



\end{abstract}



\maketitle



\section{Introduction}

Autonomous aircraft in shared airspace must navigate safely and efficiently while adhering to social norms expected in human-centric environments~\cite{gao2021evaluation}. These norms include respecting personal space~\cite{althaus2004navigation}, maintaining comfortable velocities and accelerations~\cite{kato2015may}, and keeping a safe distance from other aircraft~\cite{glozman2021vision}. Research into socially aware navigation aims to improve interactions between autonomous agents and humans~\cite{kruse2013human}, but there is still a need for new metrics to evaluate these methods more effectively~\cite{wang2022metrics}.

The domain of air combat introduces additional layers of complexity to social navigation~\cite{birkeland2018concept}. Integrating a loyal wingman with human pilots requires not only safety and efficiency but also a deep understanding of tactics and formation dynamics, demanding a sophisticated mix of social and combat skills~\cite{dantas2023autonomous}. This paper adapts socially aware navigation for air combat by introducing tailored social navigation metrics for autonomous wingmen.

Our main contribution is developing these metrics and proposing a validation process through a user study experiment with military pilots in high-fidelity simulations. This research addresses a gap in the existing literature and sets the stage for future integration of autonomous systems in manned military operations, enhancing both effectiveness and social compatibility.
\vspace{-0.1cm}
\section{Proposed Social Navigation Metrics}


To assess autonomous agents in air combat, we propose key metrics focused on naturalness and comfort. Table~\ref{tab:1} outlines these metrics, detailing the aspects evaluated and the rationale for each.


\begin{table}[thpb]
\scriptsize
\caption{Summary of the proposed social navigation metrics.}
\label{tab:1}
\begin{tabular}{|C{0.35cm}|C{1cm}|C{1cm}|L{4.7cm}|}
\hline
\makecell{\textbf{No.}} & \makecell{\textbf{Aspect}} & \makecell{\textbf{Metric}} & \textbf{Description} \\
\hline
\makecell{$M_1$}   & \makecell{Naturalness} & \makecell{Velocity} & Computes the mean of the squared velocities over the time period, highlighting significant speed variations from typical human norms\\
\hline
\makecell{$M_2$}  &  \makecell{Naturalness} & \makecell{Acceleration}   & Calculates the average of squared accelerations to assess how naturally the acceleration changes compare to human-like movements \\
\hline
\makecell{$M_3$} & \makecell{Naturalness}  & \makecell{Jerk}           & Evaluates the mean squared jerk to identify abrupt changes in acceleration, aiming for smoother, more human-like trajectories \\
\hline
\makecell{$M_4$}   & \makecell{Comfort}                      & \makecell{Minimum \\Distance} & Calculates the smallest distance between two agents by iteratively comparing their positions over a given time period and updating the minimum found \\
\hline
\makecell{$M_5$}  & \makecell{Comfort}                       & \makecell{Collision \\Risk} & Assesses the collision risk by determining how often two agents come within a critical distance or have a closing velocity that predicts a potential collision 
\\
\hline             
\end{tabular}
\end{table}

\vspace{-0.2cm}
\textbf{Naturalness:} This metric evaluates the similarity of the wingman's motion to human movements and the smoothness of its path~\cite{kruse2013human}. It involves analyzing the agent's velocity, acceleration, and jerk to assess movement smoothness and human-likeness, crucial for human trajectory prediction research~\cite{rudenko2020human}. Humans typically exhibit trajectories with compatible velocities, accelerations, and minimal jerk~\cite{kruse2013human,wang2022metrics}. To measure the wingman's trajectory smoothness, we calculate its average velocity, acceleration, and jerk to determine if these averages meet predefined thresholds that approximate human pilot levels, which depend on the type of aircraft. The squared derivatives ensure non-negativity, highlight significant variations, and smooth noise for easier mathematical handling




Refer to Equation~\ref{Eq:1} for the naturalness metrics calculation. In the equation, $p$ denotes position, with $w$ and $h$ representing wingman and human, respectively. Superscripts $w$ or $h$ indicate affiliation, subscript $t$ denotes the current time, and $T$ is the total episode duration. The symbol $\textit{n}$ indicates the derivative order, where $\textit{n}=1, 2, 3$ for velocity, acceleration, and jerk, respectively.
\begin{equation}
\scriptsize
\label{Eq:1}
M_n=\frac{1}{T}\sum_{t=0}^{T}{\left(\frac{d^n p(t)}{dt^n}\right)^2}, \quad \text{where } n = 
\begin{cases} 
1 & \text{for velocity,} \\
2 & \text{for acceleration,} \\
3 & \text{for jerk.}
\end{cases}
\end{equation}


\textbf{Comfort:} This metric assesses human comfort by minimizing disturbance in interactions with autonomous agents. It emphasizes maintaining safe distances and respecting personal spaces to reduce impact on human activities~\cite{kruse2013human}. We propose two metrics to evaluate comfort in shared airspace within air combat scenarios.

The first comfort metric measures the smallest distance maintained between the human and the wingman throughout the air combat simulation. For specific calculation details, see Algorithm~\ref{alg:1}.



\vspace{-0.2cm}

\begin{algorithm}[htp]
\scriptsize
\caption{Calculate \( M_4 \): Minimum Distance Comfort Metric}
\begin{algorithmic}[1]
\State Initialize minimum distance comfort metric: \( M_4 \gets +\infty \)
\For{\( t = 0 \) to \( T \)}
\State Calculate distance for frame \( t \): \( d_t \gets \left\| \mathbf{p}_t^w - \mathbf{p}_t^h \right\| \)
\If{\( d_t < M_4 \)}
\State Update minimum distance: \( M_4 \gets d_t \) \Comment{Record new minimum across all frames}
\EndIf
\EndFor
\end{algorithmic}
\label{alg:1}
\end{algorithm}

\vspace{-0.2cm}

\begin{figure}[thp]
\centering
\includegraphics[width=0.26\textwidth, trim={1cm 7cm 0.9cm 2.3cm}, clip]{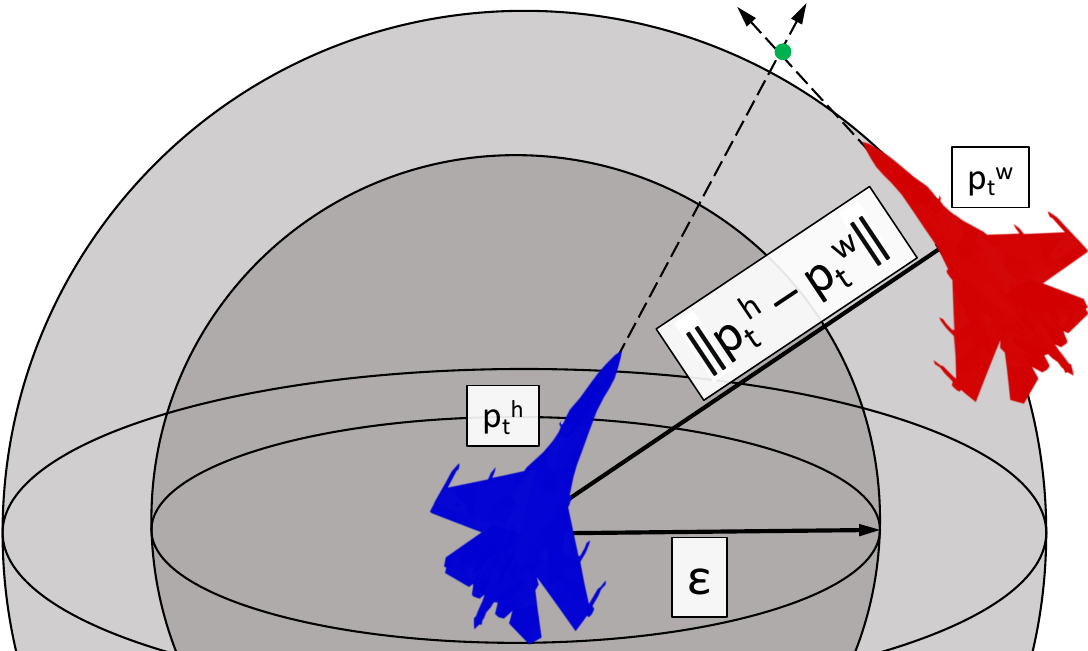}
\caption{Human-wingman personal space metric depiction.}
\label{fig:7}
\end{figure}

The second comfort metric assesses the safety of autonomous aircraft operations by calculating the risk of collisions based on the principles of the Time to Closest Point of Approach (TCPA)~\cite{glozman2021vision}. The collision risk comfort metric, \(M_5\), increments in situations where the distance \(d\) between an autonomous aircraft and a wingman is less than a critical threshold \(\varepsilon\), recommended to be set at 0.5 nautical miles for initial trials~\cite{doshi2000safe}, or when the closing velocity \(v_{\text{close}, t}\) indicates a decreasing distance that could lead to a collision within a critical time frame \(t_{\text{critical}}\). This metric effectively integrates both proximity and Time to Reach (TTR), which calculates the time until a potential collision by dividing the distance by the closing velocity, providing a comprehensive evaluation of collision risks. Refer to Figure~\ref{fig:7} and Algorithm~\ref{alg:2} for implementation details.

\vspace{-0.2cm}

\begin{algorithm}[thp]
\scriptsize
\caption{Calculate \( M_5 \): Collision Risk Comfort Metric}
\begin{algorithmic}[1]
\State Initialize collision risk comfort metric: \( M_5 \gets 0 \)
\For{\( t = 1 \) to \( T \)}
    \State Calculate relative position vector: \( \mathbf{r}_t \gets \mathbf{p}_t^h - \mathbf{p}_t^w \)
    \State Compute distance: \( d_t \gets \| \mathbf{r}_t \| \)
    \State Compute relative velocity vector: \( \mathbf{v}_t \gets \frac{d\mathbf{p}_t^h}{dt} - \frac{d\mathbf{p}_t^w}{dt} \)
    \State Calculate closing velocity: \( v_{\text{close}, t} \gets \frac{\mathbf{r}_t \cdot \mathbf{v}_t}{d_t} \)
    \If{\( d_t < \varepsilon \)} \Comment{If within critical distance}
        \State Increment collision risk comfort metric: \( M_5 \gets M_5 + 1 \) \Comment{Log alert}
    \Else
        \If{\( v_{\text{close}, t} > 0 \)} \Comment{If distance decreasing}
            \State Calculate Time to Reach (TTR): \( TTR \gets \frac{d_t}{v_{\text{close}, t}} \)
            \If{\( TTR < t_{\text{critical}} \)} \Comment{If below critical time}
                \State Increment collision risk comfort metric: \( M_5 \gets M_5 + 1 \) \Comment{Log alert}
            \EndIf
        \EndIf
    \EndIf
\EndFor
\end{algorithmic}
\label{alg:2}
\end{algorithm}

\vspace{-0.2cm}

\vspace{-0.3cm}
\section{User Study Experiment}

This user study aims to validate social navigation metrics by comparing them against human pilot perceptions in simulated air combat scenarios. The experiment will involve military pilots with varied experience, selected based on flight hours, system proficiency, and simulation experience. Experiments will be conducted in a high-fidelity simulation framework, \emph{Ambiente de Simulação Aeroespacial (ASA)}~\cite{dantas2022asa, dantas2023asasimaas}, that mimics air combat dynamics, where pilots will operate alongside a loyal wingman following the evaluated metrics. 


The simulation scenario evaluates the feasibility and effectiveness of continuous Combat Air Patrol (CAP) operations, aiming to defend a strategic point of interest, and the Defensive Counter Air (DCA) index will be used to evaluate the performance of the human-autonomous team in achieving the mission objectives~\cite{dantas2021engagement}. Data will be gathered via post-trial questionnaires assessing naturalness and comfort. The analysis will be conducted using the AsaPy Library~\cite{dantas2023asapy} to correlate these data types, validating the social navigation metrics and demonstrating their applicability in human-autonomous air combat teaming.

\vspace{-0.2cm}

\vspace{-0.3cm}

\section{Conclusion}

This study introduces social navigation metrics to enhance human-autonomous collaboration in air combat, aiming to align with pilot expectations and improve team performance. These metrics can also optimize autonomous agents' algorithms, including those based on behavior trees and reinforcement learning techniques.




\vspace{-0.1cm}

\bibliographystyle{ACM-Reference-Format}
\bibliography{ref}


\begin{thebibliography}{14}


\ifx \showCODEN    \undefined \def \showCODEN     #1{\unskip}     \fi
\ifx \showDOI      \undefined \def \showDOI       #1{#1}\fi
\ifx \showISBNx    \undefined \def \showISBNx     #1{\unskip}     \fi
\ifx \showISBNxiii \undefined \def \showISBNxiii  #1{\unskip}     \fi
\ifx \showISSN     \undefined \def \showISSN      #1{\unskip}     \fi
\ifx \showLCCN     \undefined \def \showLCCN      #1{\unskip}     \fi
\ifx \shownote     \undefined \def \shownote      #1{#1}          \fi
\ifx \showarticletitle \undefined \def \showarticletitle #1{#1}   \fi
\ifx \showURL      \undefined \def \showURL       {\relax}        \fi
\providecommand\bibfield[2]{#2}
\providecommand\bibinfo[2]{#2}
\providecommand\natexlab[1]{#1}
\providecommand\showeprint[2][]{arXiv:#2}

\bibitem[Althaus et~al\mbox{.}(2004)]%
        {althaus2004navigation}
\bibfield{author}{\bibinfo{person}{P. Althaus}, \bibinfo{person}{H. Ishiguro},
  \bibinfo{person}{T. Kanda}, \bibinfo{person}{T. Miyashita}, {and}
  \bibinfo{person}{H.I. Christensen}.} \bibinfo{year}{2004}\natexlab{}.
\newblock \showarticletitle{Navigation for human-robot interaction tasks}. In
  \bibinfo{booktitle}{\emph{IEEE International Conference on Robotics and
  Automation, 2004. Proceedings. ICRA '04. 2004}}, Vol.~\bibinfo{volume}{2}.
  \bibinfo{pages}{1894--1900 Vol.2}.
\newblock
\urldef\tempurl%
\url{https://doi.org/10.1109/ROBOT.2004.1308100}
\showDOI{\tempurl}


\bibitem[Birkeland(2018)]%
        {birkeland2018concept}
\bibfield{author}{\bibinfo{person}{John~O Birkeland}.}
  \bibinfo{year}{2018}\natexlab{}.
\newblock \showarticletitle{The Concept of Autonomy and the Changing Character
  of War}.
\newblock \bibinfo{journal}{\emph{Oslo Law Review}} \bibinfo{volume}{5},
  \bibinfo{number}{2} (\bibinfo{year}{2018}), \bibinfo{pages}{73--88}.
\newblock


\bibitem[Dantas et~al\mbox{.}(2021)]%
        {dantas2021engagement}
\bibfield{author}{\bibinfo{person}{Joao P.~A. Dantas},
  \bibinfo{person}{Andre~N. Costa}, \bibinfo{person}{Diego Geraldo},
  \bibinfo{person}{Marcos R. O.~A. Maximo}, {and} \bibinfo{person}{Takashi
  Yoneyama}.} \bibinfo{year}{2021}\natexlab{}.
\newblock \showarticletitle{{Engagement Decision Support for Beyond Visual
  Range Air Combat}}. In \bibinfo{booktitle}{\emph{2021 Latin American Robotics
  Symposium (LARS)}}. \bibinfo{pages}{96--101}.
\newblock
\urldef\tempurl%
\url{https://doi.org/10.1109/LARS/SBR/WRE54079.2021.9605380}
\showDOI{\tempurl}


\bibitem[Dantas et~al\mbox{.}(2022)]%
        {dantas2022asa}
\bibfield{author}{\bibinfo{person}{Joao P.~A. Dantas},
  \bibinfo{person}{Andre~N. Costa}, \bibinfo{person}{Vitor C.~F. Gomes},
  \bibinfo{person}{Andre~R. Kuroswiski}, \bibinfo{person}{Felipe L.~L.
  Medeiros}, {and} \bibinfo{person}{Diego Geraldo}.}
  \bibinfo{year}{2022}\natexlab{}.
\newblock \bibinfo{title}{{ASA: A Simulation Environment for Evaluating
  Military Operational Scenarios}}.
\newblock
\newblock
\showeprint[arxiv]{2207.12084}~[cs.DC]


\bibitem[Dantas et~al\mbox{.}(2023a)]%
        {dantas2023asasimaas}
\bibfield{author}{\bibinfo{person}{Joao P.~A. Dantas}, \bibinfo{person}{Diego
  Geraldo}, \bibinfo{person}{Andre~N. Costa}, \bibinfo{person}{Marcos R. O.~A.
  Maximo}, {and} \bibinfo{person}{Takashi Yoneyama}.}
  \bibinfo{year}{2023}\natexlab{a}.
\newblock \showarticletitle{{ASA-SimaaS: Advancing Digital Transformation
  through Simulation Services in the Brazilian Air Force}}. In
  \bibinfo{booktitle}{\emph{Simp\'{o}sio de Aplica\c{c}\~{o}es Operacionais em
  \'{A}reas de Defesa (SIGE2023)}} (2023-09-26). \bibinfo{pages}{6}.
\newblock
\showISSN{1983 7402}
\urldef\tempurl%
\url{https://www.sige.ita.br/edicoes-anteriores/2023/st/235455_1.pdf}
\showURL{%
\tempurl}


\bibitem[Dantas et~al\mbox{.}(2023b)]%
        {dantas2023autonomous}
\bibfield{author}{\bibinfo{person}{Joao P.~A. Dantas}, \bibinfo{person}{Marcos
  R. O.~A. Maximo}, {and} \bibinfo{person}{Takashi Yoneyama}.}
  \bibinfo{year}{2023}\natexlab{b}.
\newblock \showarticletitle{{Autonomous Agent for Beyond Visual Range Air
  Combat: A Deep Reinforcement Learning Approach}}. In
  \bibinfo{booktitle}{\emph{Proceedings of the 2023 ACM SIGSIM Conference on
  Principles of Advanced Discrete Simulation}} (Orlando, FL, USA)
  \emph{(\bibinfo{series}{SIGSIM-PADS'23})}. \bibinfo{publisher}{ACM},
  \bibinfo{address}{Orlando, FL, USA}.
\newblock
\showISBNx{979-8-4007-0030}
\urldef\tempurl%
\url{https://doi.org/10.1145/3573900.3593631}
\showDOI{\tempurl}


\bibitem[Dantas et~al\mbox{.}(2023c)]%
        {dantas2023asapy}
\bibfield{author}{\bibinfo{person}{Joao P.~A. Dantas},
  \bibinfo{person}{Samara~R. Silva}, \bibinfo{person}{Vitor C.~F. Gomes},
  \bibinfo{person}{Andre~N. Costa}, \bibinfo{person}{Adrisson~R. Samersla},
  \bibinfo{person}{Diego Geraldo}, \bibinfo{person}{Marcos R. O.~A. Maximo},
  {and} \bibinfo{person}{Takashi Yoneyama}.} \bibinfo{year}{2023}\natexlab{c}.
\newblock \bibinfo{title}{AsaPy: A Python Library for Aerospace Simulation
  Analysis}.
\newblock
\newblock
\showeprint[arxiv]{2310.00001}~[cs.MS]


\bibitem[Doshi et~al\mbox{.}(2000)]%
        {doshi2000safe}
\bibfield{author}{\bibinfo{person}{F. Doshi}, \bibinfo{person}{R. Lessem},
  {and} \bibinfo{person}{D. Mooney}.} \bibinfo{year}{2000}\natexlab{}.
\newblock \showarticletitle{The Safe Distance Between Airplanes and the
  Complexity of an Airspace Sector}.
\newblock \bibinfo{journal}{\emph{The UMAP Journal}} \bibinfo{volume}{21},
  \bibinfo{number}{3} (\bibinfo{year}{2000}), \bibinfo{pages}{257--67}.
\newblock


\bibitem[Gao and Huang(2021)]%
        {gao2021evaluation}
\bibfield{author}{\bibinfo{person}{Yuxiang Gao} {and}
  \bibinfo{person}{Chien-Ming Huang}.} \bibinfo{year}{2021}\natexlab{}.
\newblock \showarticletitle{Evaluation of Socially-Aware Robot Navigation}.
\newblock \bibinfo{journal}{\emph{Frontiers in Robotics and AI}}
  (\bibinfo{year}{2021}), \bibinfo{pages}{420}.
\newblock


\bibitem[Glozman et~al\mbox{.}(2021)]%
        {glozman2021vision}
\bibfield{author}{\bibinfo{person}{Tanya Glozman}, \bibinfo{person}{Anthony
  Narkawicz}, \bibinfo{person}{Ishay Kamon}, \bibinfo{person}{Franco Callari},
  {and} \bibinfo{person}{Amir Navot}.} \bibinfo{year}{2021}\natexlab{}.
\newblock \showarticletitle{A Vision-based Solution to Estimating Time to
  Closest Point of Approach for Sense and Avoid}. In
  \bibinfo{booktitle}{\emph{{AIAA SciTech 2021 Forum}}}. \bibinfo{pages}{0450}.
\newblock


\bibitem[Kato et~al\mbox{.}(2015)]%
        {kato2015may}
\bibfield{author}{\bibinfo{person}{Y. Kato}, \bibinfo{person}{T. Kanda}, {and}
  \bibinfo{person}{H. Ishiguro}.} \bibinfo{year}{2015}\natexlab{}.
\newblock \showarticletitle{May I Help You? Design of Human-like Polite
  Approaching Behavior}. In \bibinfo{booktitle}{\emph{Proceedings of the Tenth
  Annual ACM/IEEE International Conference on Human-Robot Interaction}}
  (Portland, Oregon, USA) \emph{(\bibinfo{series}{HRI '15})}.
  \bibinfo{publisher}{ACM}, \bibinfo{address}{New York, USA},
  \bibinfo{pages}{35–42}.
\newblock
\showISBNx{9781450328838}
\urldef\tempurl%
\url{https://doi.org/10.1145/2696454.2696463}
\showDOI{\tempurl}


\bibitem[Kruse et~al\mbox{.}(2013)]%
        {kruse2013human}
\bibfield{author}{\bibinfo{person}{T. Kruse}, \bibinfo{person}{A.~K. Pandey},
  \bibinfo{person}{R. Alami}, {and} \bibinfo{person}{A. Kirsch}.}
  \bibinfo{year}{2013}\natexlab{}.
\newblock \showarticletitle{Human-aware robot navigation: A survey}.
\newblock \bibinfo{journal}{\emph{Robotics and Autonomous Systems}}
  \bibinfo{volume}{61}, \bibinfo{number}{12} (\bibinfo{year}{2013}),
  \bibinfo{pages}{1726--1743}.
\newblock


\bibitem[Rudenko et~al\mbox{.}(2020)]%
        {rudenko2020human}
\bibfield{author}{\bibinfo{person}{Andrey Rudenko}, \bibinfo{person}{Luigi
  Palmieri}, \bibinfo{person}{Michael Herman}, \bibinfo{person}{Kris~M Kitani},
  \bibinfo{person}{Dariu~M Gavrila}, {and} \bibinfo{person}{Kai~O Arras}.}
  \bibinfo{year}{2020}\natexlab{}.
\newblock \showarticletitle{{Human Motion Trajectory Prediction: A Survey}}.
\newblock \bibinfo{journal}{\emph{The International Journal of Robotics
  Research}} \bibinfo{volume}{39}, \bibinfo{number}{8} (\bibinfo{year}{2020}),
  \bibinfo{pages}{895--935}.
\newblock


\bibitem[Wang et~al\mbox{.}(2022)]%
        {wang2022metrics}
\bibfield{author}{\bibinfo{person}{Junxian Wang}, \bibinfo{person}{Wesley~P
  Chan}, \bibinfo{person}{Pamela Carreno-Medrano}, \bibinfo{person}{Akansel
  Cosgun}, {and} \bibinfo{person}{Elizabeth Croft}.}
  \bibinfo{year}{2022}\natexlab{}.
\newblock \showarticletitle{{Metrics for Evaluating Social Conformity of Crowd
  Navigation Algorithms}}.
\newblock \bibinfo{journal}{\emph{arXiv preprint arXiv:2202.01045}}
  (\bibinfo{year}{2022}).
\newblock


\end{thebibliography}

\end{document}